\documentstyle
[psfig,epsf]
{l-aa}
\newcommand{\uSiO}     {\mbox{$^{28}$SiO$\;$}}
\newcommand{\dSiO}     {\mbox{$^{29}$SiO$\;$}}
\newcommand{\tSiO}     {\mbox{$^{30}$SiO$\;$}}

\def\,{\thinspace}
\def\etal{et al.}

\def\kms{km\thinspace s$^{-1}$}

\def\12CO{$^{12}$CO}

\begin{document}
\title{$^{29}$SiO ($v=0$) and $^{28}$SiO ($v=1$) $J=2-1$ maser emission from
Orion IRc2}

\author{A. Baudry\inst{1}, F. Herpin\inst{1} and R. Lucas\inst{2}}
\offprints{A. Baudry \\ baudry@observ.u-bordeaux.fr}
\institute {Observatoire de l'Universit\'e de Bordeaux, BP 89, F-33270
Floirac, France
\and
Institut de Radioastronomie Millim\'etrique, 300 rue de la Piscine,
F-38406 Saint Martin d'H\`eres, France
}

\thesaurus{02.12.3, 02.13.3, 08.06.2, 13.19.3}

\date{submitted 11 December 1997; accepted 31 March 1998}
\maketitle
\markboth{A. Baudry, F. Herpin and R. Lucas: \dSiO and \uSiO emission
from Orion IRc2}{}
\begin{abstract}
We have observed with the IRAM interferometer at two
different epochs and simultaneously the two
transitions $v=0, J=2 \rightarrow 1$ and $v=1, J=2 \rightarrow 1$ of
\dSiO and \uSiO in Orion IRc2. We have made the first
maps of \dSiO $v=0, J=2 \rightarrow 1$ emission from Orion. These
maps and properties of the \dSiO spectra attest to maser emission. Our
\uSiO maps show the stable ring of maser spots observed in
previous works. Combining our own data with published works we
derive that the relative motion between the two ridges of the \uSiO
emission ring
is less than about 0.7 AU/yr over a period of 7 years. On the other hand,
the weak high velocity
maser features observed around 30 \kms ~seem to move with respect
to the stable ring of \uSiO main emission. Our relative \dSiO ($v=0$) and
\uSiO ($v=1$) spot maps show that most \dSiO and \uSiO emission
features are closely
related but have not the same spatial extent. We
conclude that these masers are not excited in
the same gas layers in agreement with pumping models which predict
that various $v$ state masers peak in different spatial regions.
In addition, our maps of $v=0$ and $v=1$ emission suggest that local line
overlaps due to turbulence and high gas temperature do not play a dominant
role in the excitation of \uSiO and \dSiO, although excitation effects
resulting
from the overlap of Doppler-shifted ro-vibrational lines may
still be significant.
\keywords{ISM: Orion A; star formation; interferometry; masers: \dSiO, \uSiO}
\end{abstract}

\section{Introduction}

Since the discovery nearly 25 years ago of the SiO molecule in Orion
(Snyder \& Buhl 1974), SiO masers have
been observed in the envelopes of hundreds of late-type stars, and in the
direction of a few
galactic HII regions. Orion remains a unique source of SiO emission
because it contains all known SiO isotopic species and because it is
the only star-forming region exhibiting a very strong maser in the
$v=1, J=2 \rightarrow 1$ transition. The strong and
compact $v=1$ SiO maser is associated with the luminous infrared
source IRc2 and is closely connected with the extended and weaker
$v=0$ maser emission mapped at 43 GHz (Chandler \& De Pree 1995) and 86
GHz (Wright et al. 1995).
Recent high resolution observations showed that IRc2 is a complex
object resolved into
four components (Dougados et al. 1993) and that the center of the
strong ($v=1$) and weak ($v=0$) SiO
maser outflows coincide with the radio continuum source I
(Menten \& Reid 1995, Wright et al. 1995), and is displaced from the
center of the molecular hot core (lying to the east of source I). On
the other hand, the center of the large-scale high velocity bipolar
outflow traced by CO lies roughly $3"$ to the north of
source I.

 Considerable efforts have been made to properly model the SiO maser
phenomenon in late-type stars. Models include radiative and/or collisional
excitation
schemes (e.g. Kwan \& Scoville 1974, Elitzur 1980,
Langer \& Watson 1984, Lockett \& Elitzur 1992 or Bujarrabal 1994). All
models share
two general characteristics:
$(i)$ they require high volumic densities of order
$10^{8}-10^{10}$ cm$^{-3}$; $(ii)$ inversion of the SiO level populations
depends on the column
density, and maser emission in higher vibrational states peaks at
higher values of the column density. On the
other hand, there are major observational facts that cannot be explained
by any of the present radiative/collisional SiO pumping schemes. In
particular, the "standard" pumping schemes fail to explain in stars
the peculiar distribution of line intensities within a given
vibrational state (e.g. Cernicharo \& Bujarrabal 1993), and fail to
explain the absence or weakness of $v=2, J=2 \rightarrow 1$ emission
from Orion and late-type stars (Olofsson et al. 1981 b, Bujarrabal
et al. 1996). In fact, line
overlaps among transitions of the isotopic species of silicon
monoxide are an important addition to radiative/collisional
pumping in stars (e.g. Gonz\'alez-Alfonso \& Cernicharo 1997), while
the line overlap between two near infrared lines of SiO and water
explains the weakness of the $v=2, J=2 \rightarrow 1$ transition
(Bujarrabal et al. 1996).

 Depending on the relative importance of property $(ii)$ above (namely
various $v$ state masers should peak in different spatial regions)
with respect to line overlap effects among nearby transitions of
SiO and isotopes the spatial distribution of various vibrational transitions
should differ or not. Therefore, high spatial resolution and sensitive
maps of SiO and isotopes should provide a test of these predictions.
With this idea in mind we have
compared the spatial distributions of two nearby transitions of
\dSiO and \uSiO toward Orion IRc2 which
contains the strongest SiO source in the sky. In Sect. 2 we
present our observations and give details of data reduction. In Sect. 3
we discuss spectral variability and present our maps of \dSiO and \uSiO
emission
from Orion. In Sect.~4 we discuss some properties of the apparent ring of
\dSiO and
\uSiO masers, the relative spatial extents of both
species and implications on their excitation. Some conclusions are
summarized in Sect. 5.

\section{Observations and data reduction}
In order to compare the relative spatial distributions of the main
(\uSiO) and rare
isotope (\dSiO) emissions we need a relative positional accuracy
better than the extent of the main isotope emission, $\approx 0.15"$.
For this kind of accuracy simultaneous observations
of two different transitions are appropriate. We thus searched for nearby
frequencies
involving rotational levels not too high in energy.
The frequencies of the $v=0, J=2 \rightarrow 1$ and $v=1, J=2 \rightarrow 1$
transitions of \dSiO and \uSiO (85.75913 and 86.24337 GHz) differ by less
than the 500 MHz instantaneous bandwidth of the
interferometer and can thus be simultaneously observed.
Observations of these two lines in Orion IRc2 were made on
August 19 and 21, 1995 and March 5, 1996 with four antennas of the IRAM array
on Plateau de Bure (see Guilloteau et al. 1992 for details) . In 1995 we used
configuration B2 with antennas on stations N17 E24 W09 W12 and spacings
covering 24 to 288 m. The synthesized beam was around $3.5" \times 1.4"$.
Higher spatial resolution was achieved in 1996 with
antennas on stations N29 E24 W20 W27 corresponding to spacings from 56 to
408 m and resulting in  $2.6" \times 1"$ resolution.
The independent correlator units of the interferometer were used as
follows. The $v=0,
J=2 \rightarrow 1$ and $v=1, J=2 \rightarrow 1$ lines of \dSiO and \uSiO
were placed in two 20 MHz sub-bands
centered at 113.75 MHz and 596.25 MHz, respectively.
The frequency separation between adjacent spectral channels was
78.1~kHz, or  0.27 km s$^{-1}$.
In 1995, for
redundancy, the
\dSiO $v=0$ line was also observed in a 40 MHz wide sub-band.
Broad band
continuum observations with three or two 160 MHz units were performed
simultaneously.
\begin{figure} [tbh]
  \begin{center}
     \epsfxsize=8.5cm
     \epsfbox{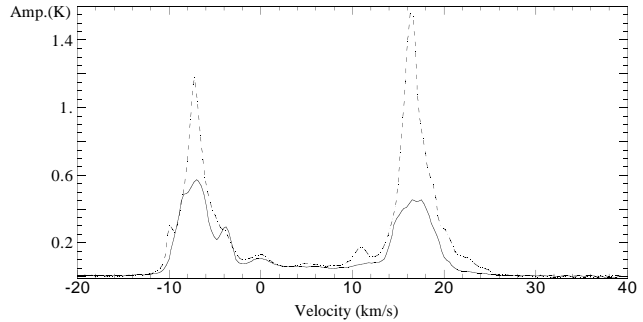}
  \end{center}
\caption[]{Cross-correlation spectrum of the $v=0, J=2 \rightarrow
1$ line of \dSiO observed with the
IRAM array on 1995 August (continuous line) and 1996 March (dotted line).
The separation between channels was
0.27 km s$^{-1}$. 1 K corresponds to a flux density of 24 Jy.}
\label{variable29SiO}
\end{figure}

 In this project we have mapped the \dSiO and \uSiO line emissions using
one of the strongest features of the main isotopic species to
self-calibrate all
data. Therefore, a careful bandpass calibration was required across the entire
bandwidth since the maximum frequency separation between extreme
features of both species is around 496 MHz. To this end we observed
strong calibrators, 3C454.3, 3C273, $0528+134$ or $0415+379$. In
1995 our best results were obtained  with
3C454.3 which we observed just before a long run in Orion. Although we do not
expect a priori any short time-scale variations of the bandpass shape, we
used a
different scheme in 1996. Observations of Orion were interspersed with
those of the bandpass calibrator $0415+379$.
Nearby phase calibrators were not essential for this project since we
self-calibrate the data on one of the main spectral features. The
bandpass calibrators were used to calibrate the flux density scale. At 86
GHz we adopted
6.5 and 10 Jy for 3C454.3 (1995 August observations) and $0415+379$
(1996 March observations), respectively. Although the flux density for
$0415+379$ was uncertain because it was highly time variable in 1995
and 1996, we averaged our measurements of the antenna temperature to Jy
conversion
factor for both observing periods and adopted S/T = 24 Jy/K.

 Data reduction was performed with the CLIC package (Lucas 1996). The sky was
clear during the observations and little editing was necessary. However,
we discarded some of the 1995 data when Orion was too low on the horizon.
The bandpass solutions
were determined by fitting the complex gains in several sub-bands ; these
solutions
were applied to the line data base. In 1995 each day
was treated separately.
\scriptsize
\begin{table} [t]
 \hspace* {-1. cm}{ {\scriptsize{
 \caption{Signal to noise ratio and errors for
 observations of 1996 March (the synthesized beam is of order 1.6").}
\begin{tabular}{l c c c}   \hline
      {\bf SNR and Error} & {\bf \uSiO}  & {\bf \dSiO} & {\bf \dSiO/ \uSiO}
\\ \hline\hline
      {\em SNR} & 200-25000 & 150-400 & 150-25000 \\
      {\em r.m.s. phase (degree)$^{1}$} & 0.002-0.29 &
      0.14-0.38 &
      0.14-0.48\\
      {\em r.m.s. position (mas)}  & 0.03-4 & 2.0-5.3 & 2.0-6.6 \\
\hline
{\em bandpass phase (degree)$^{2}$} & 1 & 1 & \\ \hline
{\em phase shift (degree)$^{3}$} & & & 1 \\
{\em position shift (mas)} & & & 4.4 \\ \hline
\end{tabular}}}}
 $^{1}$Phase noise due to finite signal to noise ratio; the
corresponding position error is ${\sigma}_{\phi}\simeq 0.5\times
1.6"/SNR$. \\
 $^{2}$Typical total bandpass phase noise; the channel to channel IF
noise is
negligible and less than 0.1 degree. \\
 $^{3}$Maximum phase shift between \uSiO and \dSiO sub-bands; one degree
phase shift corresponds to 4.4 mas with 1.6" synthesized beam.
\end{table}
\label{erreur}
\normalsize
To improve the phase stability
we determined accurate baseline solutions and we applied the known
antenna axis corrections. The phase residuals lied in the range $4
\rightarrow 30$ and $4 \rightarrow 8$ degrees for the 1995 and 1996
observations, respectively.
We then mapped the \uSiO
and \dSiO line emissions using the task UV\_ASCAL and one strong feature of
the main
isotope to self-calibrate all other features. Accordingly, we are
insensitive to atmospheric phase fluctuations and to any systematic errors
left in the baseline calibration. The position accuracy within the \uSiO
and \dSiO spot maps is limited by
finite signal to noise ratio and by imperfect bandpass calibration
(see Table 1).
In practice, the position errors within our \uSiO maps lie around
$1-2$ milliarc sec (mas) in RA and $4-5$ mas in Declination. This
allows us to discuss the long-term stability of the velocity pattern
observed for both isotopes, as well as the long-term
stability of the \uSiO emission ring discussed in Sect. 4.1.
When one compares the relative positions of the
\dSiO and \uSiO emissions, the relative phase shift of the two \uSiO
and \dSiO sub-bands cannot be neglected (Table 1); we found a maximum
drift of 1 degree per hour for the March 1996 observations. Because
we regularly calibrated the bandpass in 1996, the position error
between both isotopic species due to phase shift between sub-bands was
less than 5 mas. This error is estimated to be about
ten times higher in our data of 1995 August due to much less time
spent in bandpass calibration. In Sect. 4.4, we only discuss the
relative positions of both isotopic species observed in 1996. The
overall relative \uSiO to \dSiO position uncertainty is of order 5-8 mas.

\section{Results}

\subsection{\dSiO $v=0, J=2 \rightarrow 1$ emission}
 The \dSiO v=0 line at 85.759 GHz was
discovered in Orion by Olofsson \etal (1981 a) who observed
narrow and time variable features. The
maser nature of this emission is well demonstrated here because we detect
narrow spectral emission of same intensity with all baselines of the
array. \dSiO $v=0$ emission will be discussed further in Sect. 4.2 and
is essentially unresolved with the IRAM connected array (see below, however).
In addition, we
observe spectral variability over the 6.5 months period of our two observing
sessions since both the line shape and relative peak intensities clearly
changed with time (Fig. \ref{variable29SiO}).
The emission spectrum consists of two main features in the
velocity range $-11 \rightarrow 0$ and $10 \rightarrow 24$ km s$^{-1}$.
This is similar to the \uSiO emission range although
\dSiO and \uSiO emission profiles are much different, a fact which
can partly be related to differences in the degree of saturation of
both masers. There is weak emission in the intermediate velocity range
$0 \rightarrow 10$ km s$^{-1}$. This
\dSiO emission is about 20 times weaker than in \uSiO, and is
detected with all baselines of the array.
However, with the 24-m long baseline available in 1995,
the \dSiO emission lying between the two
main features was nearly 5 times stronger than for all other longer
baselines. We conclude that for this baseline there was a blend of
quasi-thermal and maser \dSiO emission. Both maser emission and quasi-thermal
emission were also observed in the $v=0$ state of the more abundant
species \uSiO (Chandler \& De Pree 1995, and Wright et al. 1995 for
the $J=1-0$ and $J=2-1$ transitions, respectively). Wright et al.
showed that the \uSiO $v=0$ maser component extends over $\approx 2"$
although it is closely connected with the $v=1, J=2-1$
compact masers. On the other hand, their \uSiO
$v=0$ low spatial resolution maps are more sensitive to thermal
emission and show a connection with the high
velocity bipolar CO outflow ($\approx 25" \times 45"$ in size).
For all \dSiO spectra and data discussed in
this work we restricted the analysis to $(u,v)$ distances
greater than about 50 m and 25 m for our observations of
1996 March and 1995 August, respectively. We were thus insensitive to
structures extending over $\approx 10" - 30"$.

\begin{figure} [tbh]
  \begin{center}
     \epsfxsize=8.5cm
     \epsfbox{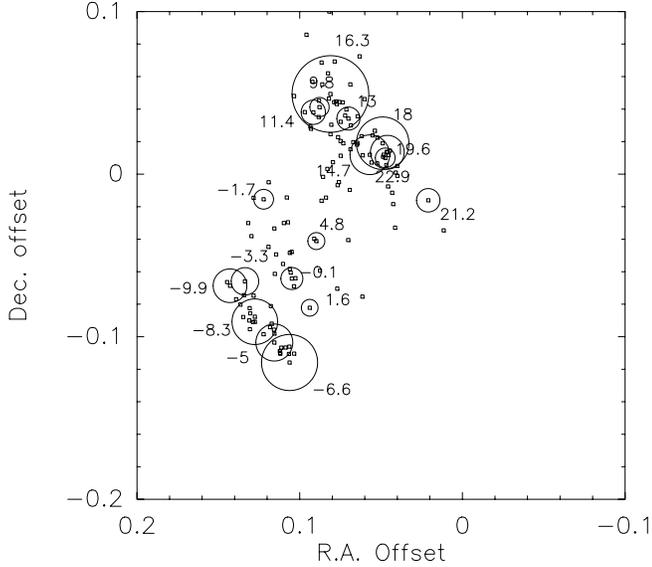}
  \end{center}
\caption[]{Spot map of \dSiO $v=0, J=2 \rightarrow 1$ emission observed on 1996
March. The small open squares correspond to the centroid of emission in one
spectral channel. Each channel is
separated by 0.27 km s$^{-1}$. The center of the brighter channels is
surrounded by a circle whose diameter is proportional to the peak
intensity in these channels. The LSR velocities are shown for the
brighter channels, and, for clarity, the circles and velocity labels are given
every 6 channels.}
\label{29SiOrelatmap}
\end{figure}

 Fig. \ref{29SiOrelatmap} shows the relative positions of \dSiO $v=0, J=2
\rightarrow 1$ maser spots observed in 1996.
We used the strong \uSiO $v=1, J=2 \rightarrow 1$ maser feature
at 15.6 km s$^{-1}$ to self-calibrate these data (cf. Sect. 2),
and we verified that selection of a specific reference feature
was not critical. The \dSiO $v=0$ emission is distributed
along two ridges of positive and negative velocities as for \uSiO
(compare Fig. \ref{29SiOrelatmap} with Figs. \ref{relat28map}
and \ref{relat28mapbis}) with stronger \dSiO $v=0$ features
excited at the edge of each ridge contrary to \uSiO. The
mean distance between the two \dSiO ridges is of order $0.13"-0.14"$.
%
This picture is fully consistent with our 1995 results
although the mean distance
between the two ridges was slightly narrower in 1995 and of order $0.11"$.
The 20
to 30 mas change between the two ridges observed in 1995 and
1996 seems real because it is greater than the typical relative position
error of the
individual \dSiO features.
At both epochs of observations the data were acquired
with non rotating antenna feeds sensitive to
vertical polarization only. However, we are not affected by polarization
effects
because we have verified that our relative spot maps
were nearly identical when we analyzed the data over short periods
with a small range of paralactic angle.
\begin{figure} [tbh]
  \begin{center}
     \epsfxsize=15.cm
     \epsfbox{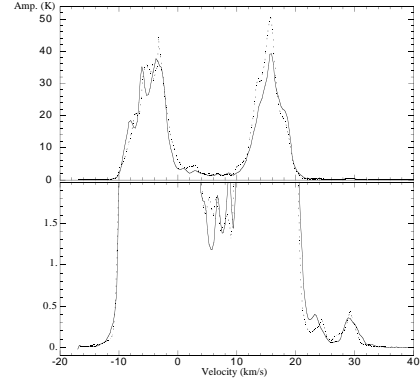}
  \end{center}
\caption[]{Cross-correlation spectrum of \uSiO $v=1, J=2 \rightarrow 1$
emission observed on 1995 August (continuous line) and 1996 March (dotted
line). The upper panel shows the overall spectrum and the lower panel
shows the weak features around 23 to 33 km
s$^{-1}$ and intermediate velocity emission around $4 \rightarrow 10$ km
s$^{-1}$. The separation between individual channels
is 0.27 km s$^{-1}$. 1 K corresponds to a flux density of 24 Jy.}
\label{weak28SiO}
\end{figure}
\subsection{\uSiO $v=1, J=2 \rightarrow 1$ emission}
The \uSiO $v=1, J=2 \rightarrow 1$ spectrum is dominated by two
strong, time variable
features lying around $-2 \rightarrow -10$ and $12 \rightarrow 20$ km
s$^{-1}$. In addition, the line profile exhibits weak
features around 30 km s$^{-1}$ which were first detected by Wright et
al. (1995; see their Fig. 1). These features are present in our 1995
and 1996 data
(Fig. \ref{weak28SiO}). Comparison with the line
profile obtained by Wright et al. with similar spectral resolution shows
that they are time variable. The peak flux around 30 km s$^{-1}$ is
around 18 Jy in Wright et al. while we measured 10 Jy in March 1996.
Variability and detectability with all six
baselines of the array suggest that these weak features are masing as
the bulk of the \uSiO emission; see, however, discussion in Sect. 4.2.
\begin{figure} [tbh]
  \begin{center}
     \epsfxsize=8.5cm
     \epsfbox{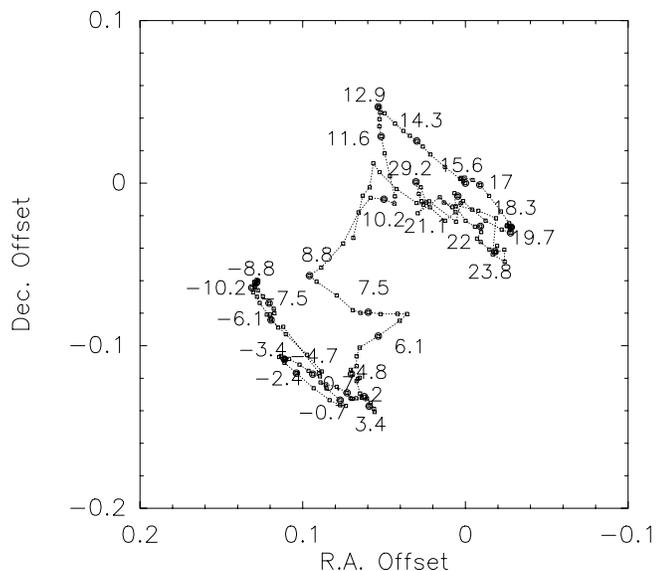}
  \end{center}
\caption[]{Spot map of \uSiO $v=1, J=2 \rightarrow 1$ emission observed
on August 19, 1995. Each velocity channel is
connected and, for clarity, the LSR velocity labels are given every 5
channels. The velocity separation between each open square is 0.27 km
s$^{-1}$.}
\label{relat28map}
\end{figure}
Our map of relative positions of $v=1, J=2 \rightarrow 1$ maser
emission observed in 1995 August is
shown in Fig. \ref{relat28map}; in Fig. \ref{relat28mapbis}
we also show the relative intensities of the main features. The
position$-$velocity pattern present in Fig. \ref{relat28map} or
\ref{relat28mapbis} is consistent with our 1996 observations (open
squares and small full circles in Fig. \ref{relat2829map}) and with
that observed in 1995 January by Wright et al. (1995).
All maps show similarities.
There are, however, minor but obvious changes among the different
maps including changes for the weak high velocity features which seem to
have moved during the period 1995 January to 1995 August and 1996 March.
On the other hand, the mean distance between the main negative and positive
velocity features remains stable. For our 1996 observations we
measured a separation of 0.16", a value very close
to that deduced from our 1995 results
and earlier IRAM results, and consistent with the BIMA array results
(see discussion in Sect. 4.1).
The absolute position of the \uSiO maser was measured by
Wright et al. (1990) and Baudry et al. (1995) to an accuracy of $\approx
0.15"-0.20$".
Menten \& Reid (1995) found that the centroid  of the SiO maser
distribution coincides with the radio continuum
source, I, lying on the southern edge of the complex source IRc2.
\begin{figure} [tbh]
  \begin{center}
     \epsfxsize=8.5cm
     \epsfbox{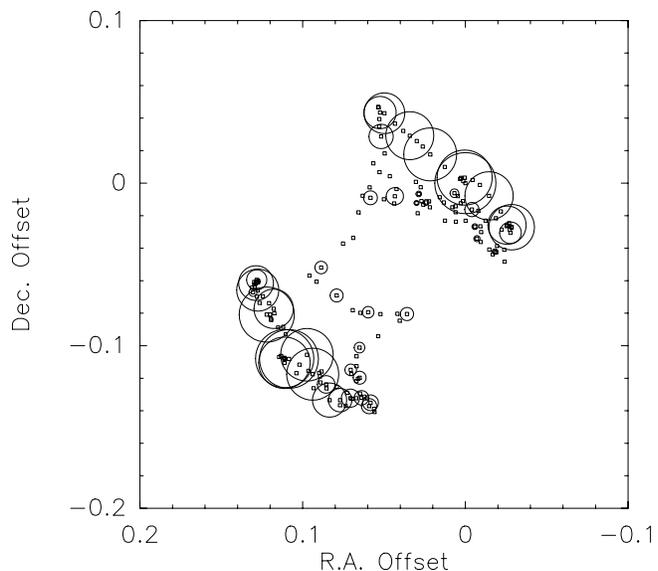}
  \end{center}
  \caption[]{Spot map of \uSiO $v=1, J=2 \rightarrow 1$ emission observed
  on August 19, 1995. Each small open square marks the center of an
  individual channel whose velocity is given in Figure \ref{relat28map}.
 The diameter of each circle, given every 3
  channels, is proportional to the line intensity.}
  \label{relat28mapbis}
\end{figure}
\section{Discussion}
\subsection{Stability of the ring of \uSiO masers}
The regular position$-$velocity pattern of $v=1, J=2 \rightarrow 1$
emission (Plambeck et al. 1990, Wright et al. 1995) was interpreted
by Plambeck et al. as a collection of
maser clumps lying in an expanding and rotating disk. This pattern
does not change much with time
and has been observed with the IRAM interferometer in 1990
(Guilloteau et al. 1992), 1992 (Baudry et al. 1995), and 1995 and
1996 (this work).
A similar position$-$velocity pattern was also observed in the $J=1
\rightarrow 0$
transition of SiO around 43 GHz by Morita \etal (1992) and Menten \& Reid
(1995).
We can estimate the long-term
stability of the 86 GHz pattern by measuring the mean separation
between the two ridges delineated by the
dominant positive and negative velocity features of
Orion IRc2. To this end, we have used the 6 different maps
made at 86 GHz with the BIMA and IRAM
interferometers (Plambeck et al. 1990, Wright et al. 1995, Guilloteau
et al. 1992, Baudry et al. 1995, and this work). The general
orientation and the mean separation between the two ridges of main SiO
features do not seem to evolve with time.
The mean separation between these two ridges is $\approx 0.^{''}165 \pm
0.^{''}01$;
for the uncertainty we have assumed that the 6 independent measurements
behaved as gaussian variables. Therefore, any apparent contraction or
expansion
of the ring, would be less than or of order $0.^{''}01 / 7$yr, namely
$\leq 0.7$ AU/yr at the 480 pc distance of Orion A. On the other
hand, stability of the
intermediate velocity pattern ($\approx 0 \rightarrow 11$ km s$^{-1}$)
is not obvious when we compare our 1990 data with the present IRAM
maps. The complex shape observed in 1995 or 1996 is not quite similar to
that in 1990 (see
Fig. 11 in Guilloteau et al. 1992). Such differences cannot be due to
relative position errors which are less than about 2 to 5 mas in the
\uSiO maps. These discrepancies seem
to agree with the model of Plambeck et al. (1990) which
predicts that intensity changes in the intermediate velocity features
could cause large position changes of the maser spots in the disk.

\subsection{Nature of the \uSiO high velocity features}

\begin{figure} [t]
  \begin{center}
     \epsfxsize=8.5cm
     \epsfbox{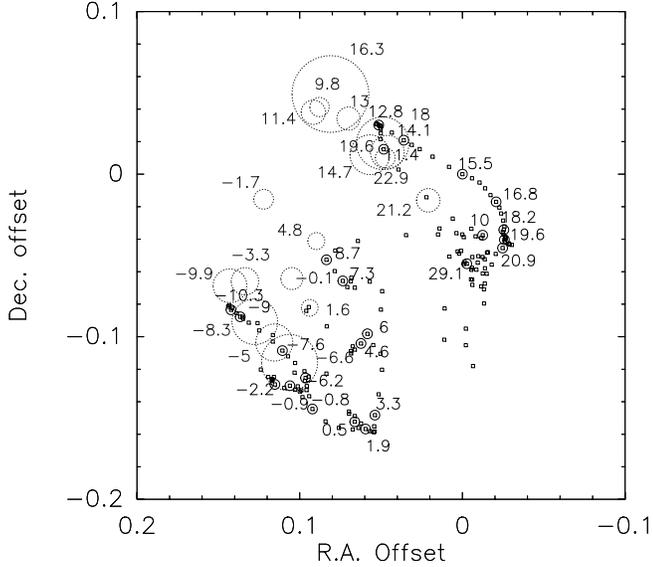}
  \end{center}
\caption[]{Comparison of \uSiO $v=1, J=2 \rightarrow 1$ and \dSiO $v=0, J=2
\rightarrow
1$ spot maps using the \uSiO feature at 15.6 km s$^{-1}$ as a phase
reference in both maps. The epoch of
the observations was March 5, 1996 for both isotopes. We have plotted the
centroids of the \uSiO features (open squares and small full circles)
together with the main \dSiO features (dotted circles). The diameter of
each circle is
proportional to the peak intensity for \dSiO. The LSR velocity labels
are given every 5 and 6 channels for \uSiO and \dSiO, respectively.}
\label{relat2829map}
\end{figure}

 The high velocity features lying around $28\rightarrow 31$ km s$^{-1}$ are
located
close to the positive
velocities of the SiO ring (see e.g. the 29.2 and 29.1 km s$^{-1}$
features in Figs. \ref{relat28map} and \ref{relat2829map}).
Our observations of 1995 August show that these features are excited in an
area similar,
although not identical, to that observed in 1995 January by Wright et al.
(1995, see their Fig. 1c)
for their $30 \rightarrow 33$ km s$^{-1}$ features.
However, 6.5 months later our data show that the $28 \rightarrow
31$ km s$^{-1}$ features have migrated toward the
most positive velocity end of the main SiO emission ridge
(see location of the 29.1 km s$^{-1}$ component in Fig.\ref{relat2829map}).
The apparent migration of weak high velocity components is consistent
with the model of Plambeck \etal (1990) where small changes in
brightness distribution of extended features may look like rapid motion.
Nevertheless, such rapid motions should be confirmed in future maps
of \uSiO emission. These high velocity
features could be related to the spectral changes observed for the
same components; they could be weakly masing as suggested in Sect.
3.2. It is interesting to note that anomalous gas motion beyond
the expansion velocity of the ring of maser
clumps could perhaps explain the high velocity components. Such components
are reminiscent of the weak features observed in the line wings of
\uSiO emission from late type stars (Cernicharo et al. 1997, Herpin et
al. 1998). SiO line wing emission in stars is related to bipolar gas
outflows and
to pulsations of the underlying star.
\begin{figure} [tbh]
  \begin{center}
     \epsfxsize=8.5cm
     \epsfbox{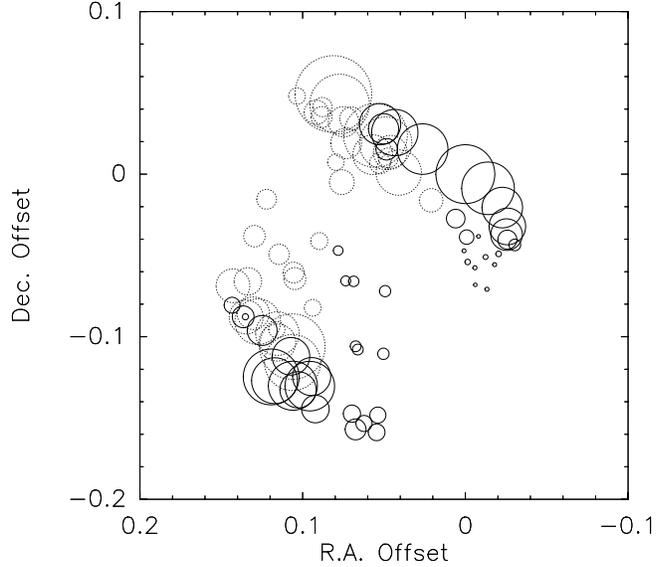}
  \end{center}
\caption[]{Comparison of \uSiO $v=1, J=2 \rightarrow 1$ and \dSiO $v=0, J=2
\rightarrow
1$ spot maps using the \uSiO feature at 15.6 km s$^{-1}$ as a phase
reference in both maps. The epoch of
the observations was March 5, 1996 for both isotopes. We only show the
centroids of
\dSiO (dotted circles) and
\uSiO (full circles) emissions. The diameter of each circle is
proportional to the peak intensity.}
\label{relat2829mapbis}
\end{figure}
\subsection{Nature of the \dSiO $v=0$ emission}
Time variability as well as notable changes in the \dSiO line profile
(Fig.
\ref{variable29SiO}) are clearly in favour of maser emission.
Short-term variability
in the excitation of the \dSiO molecule is suggested by
our maps because the mean distance observed in
1996 between the two ridges of emission, $\approx 0.13"-0.14"$, is
significantly
larger than the $0.11"$ measured in 1995. This fact indicates non
thermal processes in the excitation of \dSiO. In addition, the relatively
high flux density
observed in 1995 and 1996 suggests also non thermal emission. The
array cannot give the size of the individual features, but we may use the
synthesized beamwidth and the observed peak flux density to estimate a
minimum brightness temperature. In March 1996 the flux density peaks
around 37 Jy and we derive $T_{B} \geq 2500$ K. This temperature is greatly
above the
kinetic temperature usually adopted in Orion, $\approx 60$ K, and is
another indication
for maser emission although high temperatures would be plausible in a
shocked environment. However, for some of the weaker \dSiO features
in the range $0\rightarrow 10$ \kms we obtain T$_{B}\geq 80-100$ K,
and \dSiO could thus be part thermal and part maser.

 The actual spatial structure of the \dSiO $v=0$ emission is complex,
and we recall that Fig. \ref{29SiOrelatmap} does not show all of the
emission detectable with the array. By discarding the shorter baselines (Sect.
3.1), we have concentrated our analysis on more
compact emission sources. We were not able here to
map the \dSiO $v=0$ emission counterpart to the extended \uSiO $v=0$ emission
seen by Wright et al. (1995).

\subsection{Comparison of the \dSiO and \uSiO spot maps. Implication
on excitation mechanisms}
The relative distribution of \uSiO and \dSiO emission observed in 1996
is shown in Fig. \ref{relat2829map} and in Fig. \ref{relat2829mapbis}
where, for clarity, we have not given the velocities. The two ridges of
\dSiO $v=0, J=2
\rightarrow 1$ emission (dotted circles) and \uSiO $v=1, J=2
\rightarrow 1$ emission (full circles) are clearly visible on these
figures. Our data do not show
a complete spatial overlap as well as a close correlation among features of
both species although in both cases the positive velocities lie to the NW
of the negative features. Another clear difference between \dSiO and \uSiO
is that there is
no obvious pattern for the \dSiO intermediate velocity features (Fig.
\ref{29SiOrelatmap} or Fig. \ref{relat2829map}); this could be
related to mixed thermal and masing features as suggested in the
previous Section.

 Fig. \ref{relat2829map} shows that velocities in the range $ \approx
-6 \rightarrow -10$ km s$^{-1}$ tend to be found in the same area for both
species although it is not possible to make an exact position-velocity
pairing of
the \dSiO and \uSiO features. In Fig.
\ref{relat2829mapbis} the northern ridges of both \dSiO and \uSiO
seem to be co-aligned and are not coincident. We cannot exclude, however,
that uncorrected instrumental effects still affect our relative map and these
observations should be repeated using frequent bandpass calibrations as
used in 1996. In order to force the spatial coincidence
of the main features in both species we have shifted the main \dSiO
features lying around 16 km s$^{-1}$ on top of
the \uSiO features in the range $15 \rightarrow 17$ km s$^{-1}$. Nevertheless,
the \dSiO negative velocity emission
ridge appears well outside the \uSiO negative velocity ridge. Any rotation of
coordinates axis around the 16 km s$^{-1}$ features does not improve the
spatial coincidence of both species. Hence, we conclude that both isotopic
species
are not excited in the same gas layers. This is strengthened by the analysis
of our 1995 data which similarly show no spatial coincidence and a smaller
distance between the two ridges of emission for \dSiO than for \uSiO .

 We note that a similar picture also emerges from the 43 GHz
interferometric observations made by Morita et al. (1992) in Orion.
Although their observations of the \uSiO $v=2, J=1 \rightarrow 0$ and $v=1, J=1
\rightarrow 0$ transitions were not made simultaneously and were less
sensitive than here,
the mean separation between the two ridges of emission is slightly smaller for
$v=2$ ($\approx 0.12"$) than for $v=1$ ($\approx 0.14"$ in agreement
with the separation measured on the 43 GHz map of Menten \& Reid 1995).
The 43 GHz observations, the maps of \uSiO $v=1$ and $v=0$
emission (Wright et al. 1995) and our 86 GHz maps indicate that
different vibrational states of silicon monoxide do show a close connection
but do not exactly coincide.

 We comment below on possible explanations of the observed
similarities without exact co-location of \dSiO ($v=0$) and
\uSiO ($v=1$) maser sources. First, the silicon
monoxide reservoir seems identical for both isotopic species since
their large-scale spatial distributions are alike. This
is expected if shocks generated in the expanding flow traced by the
$v=1, J=2 \rightarrow 1$ masers enhance the sputtering of silicon
which will then react quickly in the gas phase to form both \dSiO and
\uSiO. Second, differences in the small-scale spatial distributions of
\dSiO and \uSiO could simply result from differences in the excitation
of both species or from different physical
conditions within the silicon monoxide cloud. SiO pumping models
do not require any isotopic differentiation to obtain \uSiO, \dSiO or
$^{30}$SiO maser sources. In all cases the general physical conditions
are grossly similar for one isotopic species or another apart from the
total column densities.
On the other hand, all radiative/collisional pumping models show that
different $v$ state masers peak in different spatial regions.
We believe that this fact, combined with different
degrees of saturation in the \uSiO and \dSiO masers, is essential
to explain the slightly different distribution of \dSiO ($v=0$) and
\uSiO ($v=1$) maser spots. Our maps
of relative \uSiO and \dSiO emission also show that some features from
both isotopes and with different velocities tend to be excited in
the same area. This is observed in the range $-6 \rightarrow -10$ km s$^{-1}$.
Collisional pumping
with high temperature ($\approx 1500$ K) and high molecular hydrogen density
($\approx 10^{9}-10^{10}$ cm$^{-3}$) provides
a range of \uSiO column densities where both $v=1$
and $v=2$, $J=1 \rightarrow 0$ masers are excited (Lockett \& Elitzur
1992). Such a
scheme does not apply to our apparently overlapping \dSiO ($v=0$) and \uSiO
($v=1$) 86 GHz features because their velocities are not in good agreement.
However, further observations should be conducted to investigate
the detailed kinematics and stability of the \dSiO emission.

 Line overlap effects among
various transitions of silicon monoxide cannot be ignored to explain
the excitation of this molecule.
First, {\em local} line overlaps due to turbulence play a role
as soon as the velocity dispersion reaches about 5 \kms.
Limiting ourselves to the lower $J$ values, we find that 10 to 15
ro-vibrational transitions of \uSiO, \dSiO and \tSiO overlap within 5
\kms for $\Delta v=2$ and 1. If local line overlaps would dominate the
excitation of low $J$ rotational levels in Orion, we would expect
exact spatial coincidence of the isotopic species. Our \dSiO and
\uSiO maps contradict this idea. Second, {\em non-local}
line overlap effects as described by
Gonz\'{a}lez-Alfonso \& Cernicharo (1997) in a non static
circumstellar environment are most important. The relative
distribution of \uSiO and \dSiO emission in our maps is not
inconsistent with such non-local line overlaps. In addition, it is also
plausible
that the overlap between two near infrared lines of water
and \uSiO (Olofsson et al. 1981 b) is an
important excitation process of silicon monoxide in Orion.

 Analysing the spatial extents of different $v$ state masers is clearly
important to better understand the pumping mechanisms of the SiO molecule.
This kind of work should be extended to strong stellar
SiO-emitters since the physical conditions in late-type stars and
Orion are so different. VLBI observations are required in stars in order
to make a detailed comparison of the different $v$ emission layers;
such observations have been made for the first time to map
the $v=2, v=1, J=1 \rightarrow 0$ lines in W Hya and VY CMa (Miyoshi et al.
1994).

\section{Summary}

We have made the first spot map of \dSiO $v=0, J=2\rightarrow 1$
emission from Orion, and we confirm its maser nature. However, there
is also a blend of thermal emission with maser emission.
The \dSiO $v=0, J=2\rightarrow 1$ stronger emission is
distributed along two ridges of positive and negative velocities as
for the \uSiO $v=1, J=2\rightarrow 1$ transition. The spatial extent is
similar but
not identical for the $v=0$ and
$v=1$ masers in agreement with model calculations which predict
that various $v$ state masers peak in different spatial
regions. For those apparently overlapping $v=0$ and 1
features the differences observed in the velocities suggest also that
\dSiO and \uSiO lie in different $v=0$ and 1 gas layers. Turbulence
and high gas temperature implying local line overlap effects
among transitions of \dSiO and \uSiO are unlikely to play a major role
in Orion. However, overlaps among Doppler-shifted ro-vibrational lines
are not excluded.

 We have discussed the long-term stability of
the \uSiO $v=1, J=2 \rightarrow 1$ emission
disk; relative motion between the two ridges of main emission
is less than $\approx 0.7$ AU/yr. We
confirm the detection of features at velocities as high as
about 30 km s$^{-1}$; they could be either rapidly moving weak masers
or extended features with changing brightness distribution.

\acknowledgements
{This work was supported by the CNRS URA 352. We thank the IRAM
staff on Plateau de Bure for their efficient help during the
observations, and we thank M.C.H. Wright for his useful comments.}

\end{document}